\documentclass[nonacm]{acmart}
\AtBeginDocument{%
  }

\setcopyright{none}

\usepackage[frozencache]{minted}
\usepackage{cleveref}
\usepackage{caption}
\usepackage{subcaption}
\usepackage{float}
\usepackage{graphicx}
\usepackage{hyperref}
\usepackage{siunitx}
\usepackage{booktabs}
\usepackage{multirow}
\usepackage{wrapfig}
\usepackage[inline]{enumitem}
\iffalse 
  \newcommand{\newcommenter}[3]{%
  \newcommand{#1}[1]{%
    {\color{#2}%
      \fbox{\bfseries\sffamily\scriptsize{#3}}%
      {\small$\blacktriangleright$\textsf{\emph{##1}}$\blacktriangleleft$}%
    }~%
  }%
}
\else
  \newcommand{\newcommenter}[3]{\newcommand{#1}[1]{}}
\fi

\newcommenter{\sk}{teal}{Shun}
\newcommenter{\mc}{blue}{Michael}
\newcommenter{\ak}{violet}{Ayla}
\newcommenter{\sr}{orange}{Savitha}

\begin{document}

\title{A Study of Scientific Computational Notebook Quality}

\author{Shun Kashiwa}
\email{skashiwa@ucsd.edu}
\affiliation{%
    \institution{UC San Diego}
    \city{La Jolla}
    \state{California}
    \country{USA}}
\author{Ayla Kurdak}
\email{akurdak@ucsd.edu}
\affiliation{%
    \institution{UC San Diego}
    \city{La Jolla}
    \state{California}
    \country{USA}}
\author{Savitha Ravi}
\email{s2ravi@ucsd.edu}
\affiliation{%
    \institution{UC San Diego}
    \city{La Jolla}
    \state{California}
    \country{USA}}
\author{Ridhi Srikanth}
\email{rsrikanth@ucsd.edu}
\affiliation{%
    \institution{UC San Diego}
    \city{La Jolla}
    \state{California}
    \country{USA}}
\author{Angel Thakur}
\email{athakur@ucsd.edu}
\affiliation{%
    \institution{UC San Diego}
    \city{La Jolla}
    \state{California}
    \country{USA}}
\author{Sonia Chandra}
\email{sochandra@ucsd.edu}
\affiliation{%
    \institution{UC San Diego}
    \city{La Jolla}
    \state{California}
    \country{USA}}
\author{Jonathan Truong}
\email{jmtruong@ucsd.edu}
\affiliation{%
    \institution{UC San Diego}
    \city{La Jolla}
    \state{California}
    \country{USA}}
\author{Michael Coblenz}
\email{mcoblenz@ucsd.edu}
\affiliation{%
    \institution{UC San Diego}
    \city{La Jolla}
    \state{California}
    \country{USA}}

\renewcommand{\shortauthors}{Kashiwa et al.}

\begin{abstract}
    The quality of scientific code is a critical concern for the research community.
    Poorly written code can result in irreproducible results, incorrect findings, and slower scientific progress.
    In this study, we evaluate scientific code quality across three dimensions: reproducibility, readability, and reusability.
    We curated a corpus of 518 code repositories by analyzing \textit{Code Availability} statements from all \num{1239} \textit{Nature} publications in 2024.
    To assess code quality, we employed multiple methods, including manual attempts to reproduce Jupyter notebooks, documentation reviews, and analyses of code clones and mutation patterns.
    Our results reveal major challenges in scientific code quality. Of the 19 notebooks we attempted to execute, only two were reproducible, primarily due to missing data files and dependency issues. Code duplication was also common, with 326 clone classes of at least 10 lines and three instances found among \num{637} of the \num{1510} notebooks in our corpus. These duplications frequently involved tasks such as visualization, data processing, and statistical analysis. Moreover, our mutation analysis showed that scientific notebooks often exhibit tangled state changes, complicating comprehension and reasoning.
    The prevalence of these issues---unreproducible code, widespread duplication, and tangled state management---underscores the need for improved tools and abstractions to help science build reproducible, readable and reusable software.
\end{abstract}



\maketitle

\begin{sloppypar}

    \section{Introduction}
    \label{sec:introduction}
    Software plays an increasingly important role in research \cite{johanson_software_2018}, but quality and correctness concerns \cite{trisovic_large-scale_2022,eisty_testing_2025,johanson_software_2018} and the low reproducibility rate of scientific code \cite{trisovic_large-scale_2022,tiwari_reproducibility_2021,samuel_computational_2024} threaten the validity of research findings. Motivated by these concerns, there has been a growing movement towards open science, which aims to make research more transparent, accessible, and collaborative \cite{vicente-saez_open_2018}, in part by improving the availability of research materials, including code \cite{grant_top_2025}. Jupyter, a popular environment for data analysis that combines Markdown, code, and outputs, has been lauded for supporting open science by enabling reproducible workflows, promoting shareability, and unifying code and narrative (``one study—one document'') \cite{beg_using_2021,randles_using_2017,granger_jupyter_2021}. However, Jupyter notebooks have many usability and code quality issues \cite{singer_notes_2020,pimentel_large-scale_2019,wang_better_2020,grotov_large-scale_2022}. A study of over \num{800000} general-purpose notebooks found that only 24\% were fully executable, and a mere 4\% produced the same results~\cite{pimentel_large-scale_2019}. This is troubling for scientific pursuits, where reproducibility is paramount, and may threaten the validity of findings produced by computational notebooks. Though scientific and traditional software development have distinct contexts and goals~\cite{kelly_scientific_2015,mackowiak_constructs_2018}, we can still leverage knowledge from traditional software engineering to raise scientific code quality and correctness~\cite{heaton_claims_2015,storer_bridging_2017,arvanitou_software_2021}.

    In what ways do real-world scientific Jupyter notebooks meet or fail to meet the goals of open science?
    To answer this, we collected a corpus of 518 code repositories, extracted from all \num{1239} \textit{Nature} publications in 2024. We choose \textit{Nature} because it (1) spans disciplines, so it enables examining code across many fields, each of which may have their own community standards; (2) is prestigious, so we can expect a high quality of research and associated materials; and (3) requires a \emph{Code Availability} statement in any paper where custom code was central to the results of the study.

    To clarify what it means for a notebook to ``meet the goals of open science,'' we drew on the Transparency and Openness Promotion (TOP) Guidelines~\cite{grant_top_2025}, the FAIR principles of research software~\cite{barker_introducing_2022}, and \textit{Nature}'s reporting standards~\cite{NatureReportingStandards} to identify three relevant software qualities:
    \begin{enumerate}
        \item \textit{Reproducibility:} Another researcher should be able to run the code and produce the same results.
        \item \textit{Readability:} Another researcher should be able to understand what the code does and see that it aligns with the claims of the paper.
        \item \textit{Reusability:} Another researcher should be able to adapt and build upon the code.
    \end{enumerate}
    Notably, reproducibility, readability, and reusability align with three of the code quality attributes that scientists care about when writing Jupyter notebooks~\cite{huang_how_2025}, suggesting that these qualities matter to scientists throughout the process of writing code, and not just as community standards that final code artifacts should meet.

    Our primary motivation for this work is to uncover limitations in existing computational notebooks with respect to scientific work and identify opportunities for future programming systems that better support scientists' needs. Therefore, we devised analyses to understand how specific aspects of Jupyter's design may impact reproducibility, readability, and reusability. To contextualize our findings and identify unique characteristics of scientific code, we compared scientific notebooks to general-purpose software when feasible.

    To assess reproducibility, we manually ran 19 Jupyter notebooks, following the procedures a domain scientist trying to run the code would likely use: reading documentation, downloading datasets, and troubleshooting errors. Only two notebooks were fully reproducible, and 11 of the remaining 17 failed to run due to missing data files.

    Next, we analyzed our scientific notebooks along three dimensions, chosen for their unique interaction with Jupyter features and their impact on readability and reusability: documentation, code clones, and variable mutation.

    \paragraph{Documentation}
    Jupyter notebooks are literate programming environments~\cite{granger_jupyter_2021,knuth_literate_1984} that allow end-users to interleave Markdown and code cells into a single computational narrative. The ability to communicate scientific results is part of what makes Jupyter popular among scientists~\cite{granger_jupyter_2021}. By categorizing the content of Markdown cells, we found that documentation was primarily used for section headers or to describe code behavior, but rarely added rich scientific or narrative content. Interpretation of results, discussion of reasoning, or added background knowledge each appeared in fewer than 10\% of Markdown cells. We further observed that although some notebooks were extensively documented, 29\% contained no Markdown, and scientific notebooks contained less documentation than general Github notebooks overall. Reader-facing \textit{demonstration} notebooks were generally better documented than author-facing \textit{findings} notebooks.

    \paragraph{Code clones}

    Modularity and abstraction promote readability and reusability by enabling independent understanding and reuse of components~\cite{parnas_modules}. In contrast, Jupyter notebooks organize code into cells that share a single global scope and offer no abstraction, encouraging frequent copy-paste actions~\cite{huang_how_2025}. Furthermore, functions cannot span multiple cells, which limits their explorability, since subdividing code allows users to inspect intermediate values and provides fine-grained control over organization and execution~\cite{rawn2025_pagebreaks}. This limitation may discourage users from abstracting code into functions. Additionally, notebook code cannot be exported, so to reuse it in another file, it must be copied over or extracted into a Python script. Because Jupyter's design encourages copy-paste and discourages abstraction, notebooks may be particularly prone to \textit{code clones}, segments of identical or nearly identical code. Indeed, prior work indicates that clones are prevalent in Jupyter notebooks~\cite{koenzen2020_jupyterClones,kallen_jupyter_2021}.

    Code clones can signal poor modular design, error proneness, and higher maintenance cost, harming reusability~\cite{roy_survey_2007}. However, cloning can be beneficial: clones are easy to create and adapt~\cite{kapser_cloning_2006} which is important in exploratory programming, and can be more comprehensible than abstractions~\cite{roy_survey_2007}. We found that \num{42}\% of notebooks contained a clone that was at least 10 lines long and occurred three or more times. Qualitative analysis of clone patterns revealed missed opportunities for abstraction, for example, clones that could easily be parametrized into functions. Clones that spanned multiple cells or entire files highlighted a structural limitation of Jupyter, which does not support abstraction over multiple cells.

    \paragraph{Mutation}

    Coupling, the degree to which software components are interconnected, impacts the reusability of a system and its subparts \cite{mehboob_reusability_2021}. Most traditional coupling metrics are defined at the class or component level \cite{mehboob_reusability_2021}, making them unsuitable for computational notebooks, which typically lack formal structure such as functions or classes \cite{grotov_large-scale_2022}. To address this gap, Grotov et al.~\cite{grotov_large-scale_2022} defined a metric of notebook coupling based on the number of variables shared between code cells. However, this metric captures fundamental complexities of any sequential process---data gets passed from step to step. To avoid including benign dataflow patterns in our metric, we designed a more nuanced measure focused on mutation. 

    Mutation adds complexity that makes programs harder to reason about~\cite{abelson1985,bloch2008effective,10.1109/TSE.2003.1214329}. In Jupyter, this becomes even more problematic due to any-order cell execution, which provides flexibility during development, but can lead to unreproducible state if a cell that mutates the environment is run nonsequentially. These unrestricted hidden dependencies make this form of mutation-based coupling particularly hard to reason about. Tight coupling between distant portions of a notebook suggests that large regions of code must be understood together (impairing readability) and that subparts of a notebook cannot be easily extracted (impairing reusability).

    We introduce two metrics, the \textit{mutation ratio}, which measures the number of mutating statements against total statements, and the \textit{mutation diffusion score}, which measures the distance between mutations of each variable. We found that although scientific and general-purpose notebooks have a similar amount of mutation, the mutation is more interspersed throughout scientific notebooks, suggesting greater entanglement. The mutation ratio and diffusion score were higher in notebooks than in Python scripts, both general-purpose and scientific.







    The contributions of this paper are:
    \begin{enumerate}
        \item We constructed and characterized a corpus of 518 code repositories from all \textit{Nature} papers published in 2024 (\Cref{sec:corpus}).
        \item We evaluated the reproducibility of scientific notebooks and found that only two of 19 were fully reproducible, with missing data files being the main cause of failure (\Cref{sec:reproducibility}).
        \item We explored the relationship between features of scientific notebooks and their readability and reusability through analyses of documentation (\Cref{sec:documentation}), code clones (\Cref{sec:clones}), and mutation patterns (\Cref{sec:mutation}).
        \item Based on the gaps between the code artifacts and goals, we identified needs and opportunities for future programming systems (\Cref{sec:discussion}).
    \end{enumerate}

    \section{Corpus Collection}
\label{sec:corpus}

Many journals in addition to \textit{Nature}, such as \textit{Science}, \textit{Cell}, \textit{PLOS}, and \textit{AGU}~\cite{science_standards,cell_standards,plos_standards,agu_standards}, require authors to make their code and data available at the time of publication. The code associated with these papers can be libraries, tools, or data analysis scripts, and they can be written by scientists themselves or professional research software engineers. Because of our focus on the connection between the code and the paper, we narrowed our scope to computational notebooks written by the paper's authors. In the following section, we record how many repositories included Jupyter notebooks and what programming languages the authors used.

\subsection{Corpus Collection Method}
\label{sec:corpus-collection}
We collected papers published in the journal \emph{Nature}, which publishes articles across many scientific disciplines. We analyzed all \num{1239} articles published by \emph{Nature} in 2024, finding \num{694} that included a \emph{Code Availability} section. From the text of this section, we extracted \num{1088} links, with \num{741} pointing to GitHub or Zenodo. To automate our extraction and repository download processes, we discarded links that did not point to either of these hosting sites and repositories that were larger than ten gigabytes. Some repositories were uploaded to both Zenodo and Github, and we retained the Zenodo versions; we consider these to be more archival, as they are fixed in time and receive DOIs.

To identify repositories written by the paper authors, we manually labeled repositories based on commit history and whether the code was uploaded by an author.
After this initial filtering, we manually inspected the remaining 519 repositories. We identified one repository associated with a paper that listed \num{2889} authors and contained \num{51976} files, vastly exceeding the median repository size of 14 authors and 15 files in our collection. Due to these extreme characteristics, we excluded this repository from our corpus to avoid skewing the subsequent analysis.

\subsection{Programming Languages}
\label{sec:programming-languages}
\begin{figure}[tb]
    \centering
    \includegraphics[width=\linewidth]{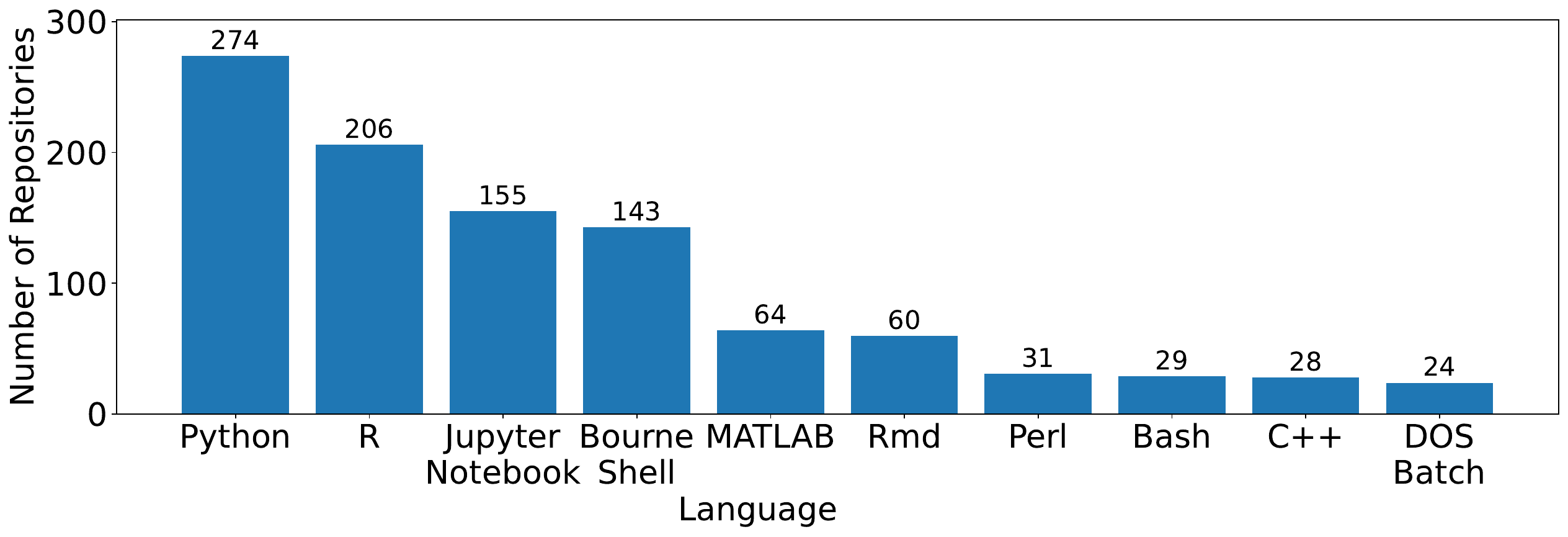}
    \caption{Top Programming Languages/Systems by Repository Count}
    \label{fig:top-prog-lang-repo-count}
    \Description{A bar chart showing the top programming languages by repository count, along with the number of repositories using Jupyter notebooks.}
\end{figure}

A scientist may choose to use a programming language based on what their lab uses, domain-specific tool interface requirements, or many other reasons. We analyzed the programming languages used in the repositories to help us understand whether the languages chosen to write research code align with languages commonly used for general software engineering.
We used the \texttt{cloc} tool~\cite{cloc} to identify the programming languages used in each repository. When analyzing some notebooks, \texttt{cloc} was unable to determine the language used, and we labelled these ``unknown,'' as was done in Pimentel et al.~\cite{pimentel_large-scale_2019} and Källén et al.~\cite{kallen_jupyter_2021}.
\Cref{fig:top-prog-lang-repo-count} shows the distribution of the top ten programming languages by number of repositories. Python was the most prevalent language, appearing in 274 repositories, followed by R with 206 repositories.

In 2011, Prabhu et al.~\cite{prabhu_survey_2011} noted that scientists often used a combination of numerical and general-purpose languages in their work. They found that languages like MATLAB and C++ were the most popular among scientists across various domains~\cite{prabhu_survey_2011}. C++ was the fourth-most popular programming language overall at the time~\cite{tiobe2011}.
Our corpus shows a similar trend, where scientists use popular general-purpose languages like Python as well as scientific, numerical language like R. By the TIOBE index, Python is the most popular language, and R is ranked 14th~\cite{tiobe2025}.

\begin{table}[b]
    \caption{Programming Languages Used in Jupyter Notebooks by File Count}
    \label{tab:jupyter-languages}
    \centering
    \begin{tabular}{@{}lrrrr@{}}
        \toprule
        \textbf{Language} & \textbf{Repositories} & \textbf{\shortstack[l]{Scientific                                       \\ Notebooks}} & \textbf{\shortstack[l]{Källén         \\ Notebooks~\cite{kallen_jupyter_2021}}} & \textbf{\shortstack[l]{Pimentel         \\ Notebooks~\cite{pimentel_large-scale_2019}}} \\
        \midrule
        Python            & 141                   & 1,434 (78\%)                      & 2,592,892 (95\%) & 1,081,702 (93\%) \\
        R                 & 15                    & 221 (12\%)                        & 21,432 (0.79\%)  & 15,204 (1.3\%)   \\
        Julia             & 2                     & 94 (5.1\%)                        & 22,336 (0.82\%)  & 10,772 (0.93\%)  \\
        Unknown           & 20                    & 82 (4.5\%)                        & 59,464 (2.2\%)   & 43,204 (3.7\%)   \\
        \bottomrule
    \end{tabular}
\end{table}

Jupyter was the most common interactive platform, appearing in 155 repositories. Many languages are supported by Jupyter notebooks. Python was dominant amongst notebooks, appearing in 141 repositories and 78.3\% of notebooks by file count. Working with datasets of general Jupyter notebooks, both Pimentel et al.~\cite{pimentel_large-scale_2019} and Källén et al.~\cite{kallen_jupyter_2021} found a majority (93\%, 95\%) are written in Python. Thus, scientists appear to use a greater variety of languages with Jupyter notebooks, although Python still dominates. \Cref{tab:jupyter-languages} shows the distribution of programming languages used in Jupyter notebooks by file count along with the findings from these two studies. \Cref{tab:nb-stats} provides summary statistics of notebook size.


In scientific repositories, Jupyter notebooks rarely appear alone, with only 15 repositories containing exclusively Jupyter notebooks. Instead, notebooks most commonly accompany Python (114 repositories), shell (53), and R (43) scripts, indicating that scientists tend to use Jupyter notebooks for specific purposes, such as analyzing data, in conjunction with code written in other languages.




\begin{table}[tb]
    \caption{Statistical summary of code metrics for Jupyter notebooks in our corpus. }
    \label{tab:nb-stats}
    \centering
    \begin{tabular}{@{}lrrrr@{}}
        \toprule
        \textbf{Statistic} & \textbf{Code Cells} & \textbf{Non-empty LoC} & \textbf{LoC} & \textbf{File Size (bytes)} \\
        \midrule
        Min                & 0                   & 0                      & 0            & 72                         \\
        25\%               & 10                  & 92                     & 104          & 25,259                     \\
        Median             & 19                  & 194                    & 224          & 153,678                    \\
        Mean               & 30.3                & 295.6                  & 347.7        & 2,194,941.5                \\
        75\%               & 35                  & 368                    & 426          & 1,281,963                  \\
        Max                & 426                 & 9,508                  & 11,636       & 79,120,237                 \\
        \bottomrule
    \end{tabular}
\end{table}


    \section{Reproducibility}
\label{sec:reproducibility}


Reproducibility is central to the credibility of computational research~\cite{tiwari_reproducibility_2021,samuel_computational_2024}---researchers should be able to rerun analyses and obtain the same outputs. Achieving this requires making code and data accessible, clearly specifying dependencies, and documenting the execution environment. In this section, we examine the extent to which the Jupyter notebooks in our corpus are reproducibile.

\subsection{Method}

To evaluate the reproducibility of Jupyter notebooks in our corpus, we manually ran notebooks from a random sample of repositories. Since some repositories contained many notebooks, we sampled a single notebook from each. If the documentation indicated file ordering, we selected the first notebook to limit potential dependencies. Otherwise, we selected a random notebook. Sampling continued iteratively until no additional error categories or reproducibility patterns emerged. We excluded repositories that required specialized hardware.

We followed guidelines from~\citet{pilgrim_ten_2023}, which include a list of steps for running other people's code and a recommendation to set up virtual environments.
We cloned each repository and created a virtual environment using Python's \texttt{venv module}. Then, we generated a \texttt{requirements.txt} file with \texttt{pipreqs}, installed the dependencies, and manually added any additional packages listed in the documentation. When setup instructions were available, we followed them; in their absence, we executed the selected Jupyter notebook sequentially from top to bottom.

Errors can arise beyond those relating to environment setup, requiring substantial effort to resolve without guarantees of success~\cite{pilgrim_ten_2023}. We spent up to one hour resolving any non-environmental errors.
To address \texttt{ModuleNotFoundError}, \texttt{ImportError}, and \texttt{RuntimeWarning}, we consulted error messages and used online resources to find potential solutions. In cases of \texttt{FileNotFoundError}, we examined the code base and the paper's \emph{Data Availability} section to locate or generate the necessary files.
If a blocking error could not be resolved within an hour, we documented the issue and proceeded to the next repository.

\subsection{Results}

Of the 19 Jupyter notebooks we evaluated, only two were fully reproducible, running fully and producing the expected results. One of these required modifying a single file path, but the documention indicated how and where to make this change. We classified an additional two artifacts as having an ``unknown'' reproducibility status. One artifact emitted a \texttt{FileNotFoundError} error, but the dataset referenced in the publication was extensive. Given enough time and effort, such as downloading the entire dataset and searching for the required file, we could likely have made the code work. However, we could not locate the file within an hour. The second artifact included a well-documented script that did not complete running within the allotted time. Still, due to its clarity and completeness, we believe it would likely reproduce successfully if given more time.

Of the 15 unreproducible notebooks, 11 encountered errors due to missing data files; the required datasets were  absent from the repository and not linked through the associated \emph{Data Availability} sections. In many cases, notebooks contained hardcoded placeholder paths without providing any guidance on how to access or obtain the necessary data. Four artifacts failed due to module import errors or incompatible dependencies. In two cases, the notebooks attempted to call functions that could not be found in the specified modules---possibly due to incorrect references or missing names. Version mismatches between libraries led to runtime failures in one notebook. The final failure occurred because a required module with a specified version number could not be downloaded, making it unavailable during execution.
These findings highlight that many scientists' practices are insufficient for ensuring the reproducibility of their artifacts due to limited dependency management, a lack of data accessibility, and gaps in documentation.


    \section{Documentation}
\label{sec:documentation}

Computational narrative, where code coexists with story, is central to the Jupyter design vision~\cite{granger_jupyter_2021}. This makes Jupyter attractive for scientific and other data-oriented work, where process, reasoning, and analysis are just as important as the code itself.
Documentation allows other scientists to understand what the code does, interpret the results of the program, and adopt the code for their own uses. While there are existing strategies and standards for documentation in software engineering contexts~\cite{aghajani2020}, scientists may have different priorities when documenting research code. We sought to understand how scientists currently document their notebooks and compare it to the documentation of general Github notebooks and high quality Kaggle notebooks.





\subsection{Method}

We were inspired by the work of~\citet{wang_documentation_2022}, who open coded the Markdown cells of 80 highly rated Kaggle notebooks. Using their nine categories as our initial set of codes, two of our authors coded the 19 notebooks sampled for our reproducibility analysis. Observing that notebooks used Markdown cells and inline comments in inconsistent and sometimes interchangeable ways, we extended~\citet{wang_documentation_2022} by also including inline comments.


\subsection{Results}

\paragraph{Summary statistics.}


Notebooks in our corpus contained a median of 3 Markdown cells and a word count of 19. This is much lower than general Github notebooks from~\citet{rule_exploration_2018}, which had a median of 218 words, despite containing less code (85 LoC compared to our 224 LoC). High quality Kaggle notebooks from~\citet{wang_documentation_2022} had a median of \num{1728} words. 29\% of notebooks in our corpus contained no Markdown, while others were extensively documented, with a maximum of \num{178} Markdown cells and \num{8456} words.


\paragraph{Markdown categorization.} Among the 19 notebooks we analyzed, there were an average of eight Markdown cells and 21 inline comments per notebook. \Cref{tab:nb-doc-categories} shows the categorization of Markdown and inline comments in scientific notebooks compared to high quality Kaggle notebooks from~\citet{wang_documentation_2022}. Headlines were the most common use of Markdown, at 47.6\%. This was followed by descriptions of code behavior (labeled as Process) with 26.9\%. These were the top two categories for the Kaggle notebooks as well, though their order was reversed (Process: 58.7\%, Headline: 32.4\%). Headlines may have greater payoff during code development than other forms of Markdown, since they delineate notebook sections, easing navigation, and their brevity makes them quick to write. Code behavior was also captured by inline comments where Process was the most common category (55.0\%), though inline comments tended to detail finer-grained information than Markdown (e.g. a single line of code).

Documentation that added context or deepened understanding was infrequent. Only 9.7\% of Markdown cells discussed the code's output, 6.9\% explained reasoning for the code, and 5.5\% supplemented the code with rich background information. These were even less common in inline comments. This suggests that although notebooks \emph{can} be used to unify computation and scientific thinking, real notebooks tend to be more code-oriented than explanation-oriented.


\begin{table}[htbp]
    \caption{Categorization of Markdown and inline comments of scientific notebooks compared to high quality Kaggle notebooks from~\citet{wang_documentation_2022}. Seven (denoted by *) are unique to our analysis relative to~\citet{wang_documentation_2022}. Percents do not sum to 100\% because Markdown and comments can have multiple categories.}
    \label{tab:nb-doc-categories}
    \resizebox{\textwidth}{!}{%
        \begin{tabular}{@{}lllll@{}}
            \toprule
            \textbf{Category}                                                                                                                                         &
            \textbf{Description}                                                                                                                                      &
            \textbf{\begin{tabular}[c]{@{}l@{}}Scientific \\ Markdown\end{tabular}}                                                                                   &
            \textbf{\begin{tabular}[c]{@{}l@{}}Scientific \\ Inline\end{tabular}}                                                                                     &
            \textbf{\begin{tabular}[c]{@{}l@{}}Kaggle \\ Markdown\end{tabular}}                                                                                                                                                                                              \\ \midrule
            Headline                                                                                                                                                  &
            \begin{tabular}[c]{@{}l@{}}The Markdown cell contains a headline. For inline \\ comments, extra characters add visual weight (e.g. \#\#\#).\end{tabular}  &
            69 (47.6\%)                                                                                                                                               &
            18 (4.4\%)                                                                                                                                                &
            1,167 (32.4\%)                                                                                                                                                                                                                                                   \\ \midrule
            Process                                                                                                                                                   & Describes what the following code is doing.            & 39 (26.9\%) & 223 (54.9\%) & 2,115 (58.7\%) \\ \midrule
            Result                                                                                                                                                    & Explains the output.                                   & 14 (9.7\%)  & 0            & 692 (19.2\%)   \\ \midrule
            \begin{tabular}[c]{@{}l@{}}Meta-\\ information\end{tabular}                                                                                               &
            \begin{tabular}[c]{@{}l@{}}Provides meta-information such as project overview, \\ author's information, and data sources.\end{tabular}                    &
            11 (7.6\%)                                                                                                                                                &
            2 (0.5\%)                                                                                                                                                 &
            141 (3.9\%)                                                                                                                                                                                                                                                      \\ \midrule
            Reason                                                                                                                                                    &
            \begin{tabular}[c]{@{}l@{}}Explains the reasons why certain functions are used \\ or why a task is performed.\end{tabular}                                &
            10 (6.9\%)                                                                                                                                                &
            22 (5.4\%)                                                                                                                                                &
            227 (6.3\%)                                                                                                                                                                                                                                                      \\ \midrule
            \begin{tabular}[c]{@{}l@{}}Background \\ knowledge\end{tabular}                                                                                           &
            \begin{tabular}[c]{@{}l@{}}Provides rich background knowledge that may not \\ be relevant to specific code.\end{tabular}                                  &
            8 (5.5\%)                                                                                                                                                 &
            8 (2.0\%)                                                                                                                                                 &
            414 (11.5\%)                                                                                                                                                                                                                                                     \\ \midrule
            Instruction*                                                                                                                                              & Instructs the reader how to use or alter the notebook. & 8 (5.5\%)   & 9 (2.2\%)    & N/A            \\ \midrule
            Assumption*                                                                                                                                               & Explains what the code assumes of its input.           & 5 (3.4\%)   & 0            & N/A            \\ \midrule
            Label*                                                                                                                                                    &
            \begin{tabular}[c]{@{}l@{}}Labels a detail of the code or data, often units or the \\ meaning of a value. For example, ``biovolume in um3.''\end{tabular} &
            4 (2.8\%)                                                                                                                                                 &
            108 (26.6\%)                                                                                                                                              &
            N/A                                                                                                                                                                                                                                                              \\ \midrule
            Reference                                                                                                                                                 & Contains an external reference.                        & 2 (1.4\%)   & 6 (1.5\%)    & 200 (5.6\%)    \\ \midrule
            Todo                                                                                                                                                      & Describes a list of actions for upcoming analysis.     & 1 (0.7\%)   & 0            & 202 (5.6\%)    \\ \midrule
            Note to self*                                                                                                                                             & Reminders, questions, or thoughts for the code author. & 1 (0.7\%)   & 9 (2.2\%)    & N/A            \\ \midrule
            Docstring*                                                                                                                                                & Docstring that describes the purpose of a function.    & N/A         & 12 (3.0\%)   & N/A            \\ \midrule
            Program flow*                                                                                                                                             & Code commented in/out to control program flow.         & N/A         & 5 (1.2\%)    & N/A            \\ \midrule
            Separator*                                                                                                                                                & Empty comment used for visual separation.              & N/A         & 3 (0.7\%)    & N/A            \\ \midrule
            Summary                                                                                                                                                   & Summarizes what has been done so far.                  & 0           & 0            & 51 (1.4\%)     \\ \bottomrule
        \end{tabular}%
    }
\end{table}

\paragraph{Notebook styles.} We subdivided our 19 notebooks into reader-facing \textit{demonstration} notebooks and author-facing \textit{findings} notebooks. Demonstration notebooks were explicitly stated as such, either through their name (e.g. containing the word "example" or "demo"), or in the Readme. Demonstration notebooks tended to be better documented than findings notebooks, introducing and contextualizing the notebook, interspersing paragraph-length explanations between code cells, and organizing the notebook with section headers. Six of the seven demonstration notebooks matched this style of documentation. In contrast, of the 12 findings notebooks, six contained no Markdown, five were minimally documented with only headlines or a few brief sentences, and only one was heavily documented. These results align with the findings of~\citet{rule_exploration_2018} who also observed that among academic notebooks, tutorial notebooks were more extensively documented than analysis notebooks.

    \section{Code Clones}
\label{sec:clones}

Code clones are segments of identical or similar code often viewed as indicators of poor design that can harm \textit{readability} (readers must track multiple similar code segments and understand their subtle differences) and \textit{reusability} (functionality is scattered across duplicated segments rather than abstracted into reusable components)~\cite{roy_survey_2007}. Jupyter's design makes notebooks particularly susceptible to code cloning due to the limited abstraction mechanisms. In this section, we analyze cloning patterns in scientific notebooks to understand their prevalence, causes, and implications.

\subsection{Clone detection}

We used NiCad~\cite{nicad}, a well-validated and widely-used clone detection tool, to identify clones across all \num{1510} notebooks and \num{4816} Python scripts. NiCad supports detection at multiple granularities, including at the level of statement sequences (block-level) and at the file level. By default, NiCad detects clones with a 70\% similarity threshold to capture clones that contain variants.
Since NiCad does not natively support Jupyter notebooks, we developed a preprocessing step to convert notebooks into analyzable Python scripts. We transformed each notebook cell into a Python function, preserving the cell's sequential structure while enabling NiCad to detect cell-level clones as block-level patterns. This approach maintains the semantic boundaries of notebook cells and allows us to detect clones across both Jupyter and Python files. 

\subsection{Cloning Patterns}


To identify patterns of clone occurrences, we performed a qualitative analysis of block-level clones. We cast a wide net, detecting clones with as few as three statements. This low threshold did produce false positives, but we manually removed these instances. NiCad identified \num{2466} clone classes (groups of similar code segments) in \num{954} of the \num{1510} notebook files (63\%) in our corpus. Of these, \num{246} (10\%) classes occurred in both a notebook and Python script, while the rest occurred only in notebooks. We randomly sampled 30 notebooks containing clones. We manually coded each clone class of these notebooks along four dimensions: the activity performed, the structure, differences between instances of the class, and the context in which instances occurred (e.g. within- or between-files). From these low-level codes, we identified five cloning patterns: boilerplate, notebook templates, identical data pipelines, duplicated function definitions, and potential functions.


\textbf{Boilerplate} comprises repetitive setup patterns that are necessary for configuration and contain minimal variation. These clones typically appear across different notebooks as standard initialization sequences. Examples include import statements, plotting configuration, data loading patterns, and model initialization. Boilerplate is also common outside notebooks and is difficult to refactor due to limited abstraction mechanisms (e.g., no macro systems or advanced templating).

\textbf{Notebook Templates} are duplicate notebooks, often performing the same analysis on new data. We discuss this type of clone more in \Cref{subsec:file-level-clones}.

\textbf{Identical Data Pipelines} are clones similar to notebook templates, but they only involve a few cells and may vary more between instances. These occur when heterogeneous datasets require customized steps of the same process. These clones can appear multiple times in the same notebook or across different notebooks. The code used in these clones could be refactored into functions, but this could result in having numerous parameters to account for edge cases when working with new data.

\textbf{Duplicated Function Definitions} are one or more function definition copied across files, instead of being imported them from a library.

\textbf{Potential Functions} are clones that have a single purpose and would require few parameters if refactored into functions. This is the broadest category of clones and involves code from a wide range of sources, including loop bodies and entire cells.

\subsection{High-Impact Block Clones}

\begin{figure}[htbp]
    \centering
    \includegraphics[width=.8\columnwidth]{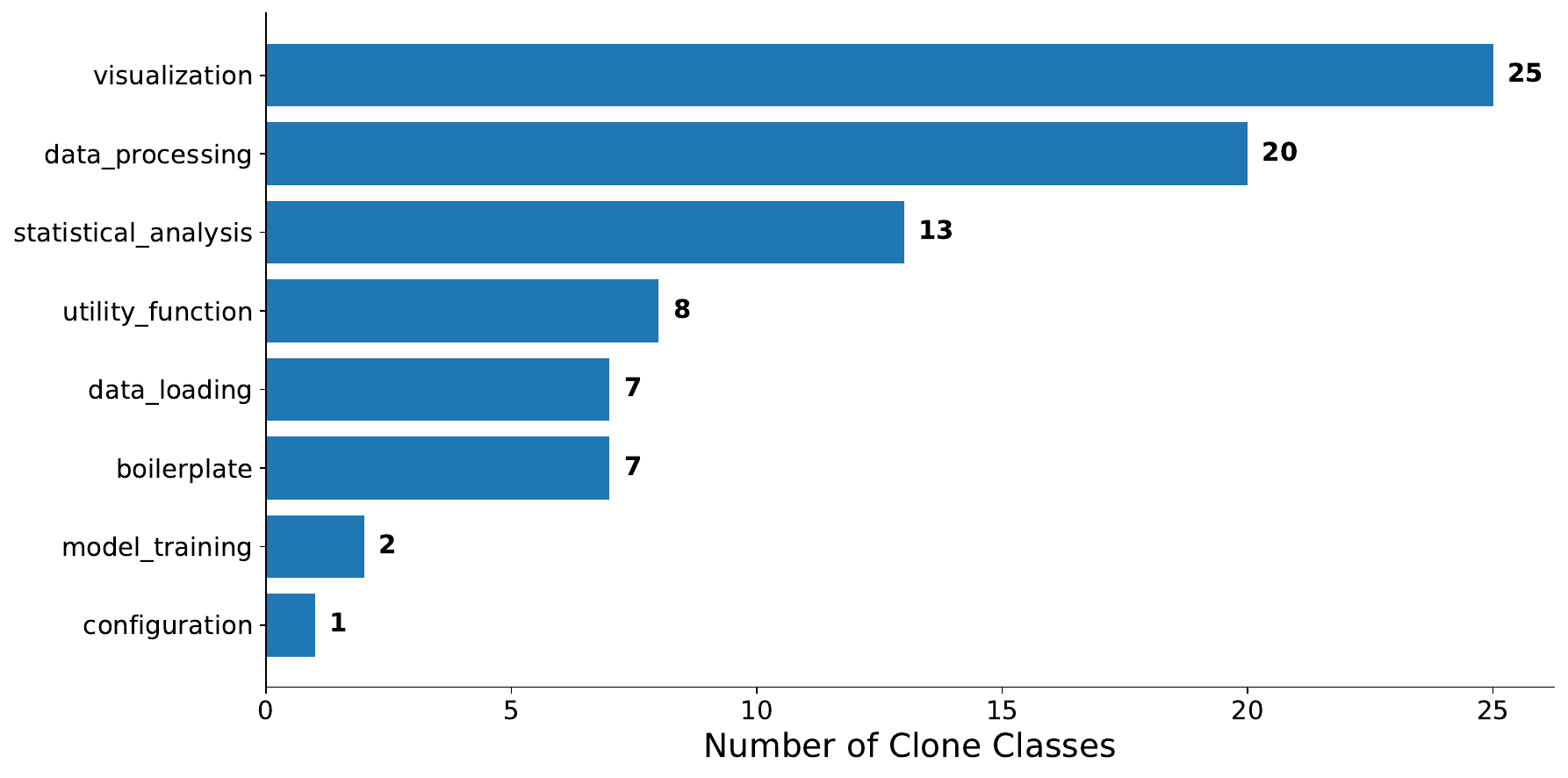}
    \caption{Categorization of high-impact block clones}
    \label{fig:clones-interesting-pattern-distribution}
\end{figure}

In our initial block-level clone analysis, we detected \num{2466} clone classes. However, \num{1247} of those clone classes only contained two instances and \num{1428} of them were under 10 lines long. While these are technically clones, they may not be particularly problematic for maintainability. To identify clones where refactoring could have greater impact on maintainability, we focused on larger, more prevalent clones, only including those with at least 10 lines of code and three occurrences. We excluded notebooks that are cloned at the file level as these are analyzed separately in the next subsection. \num{496} clone classes met these criteria. We sampled 50 of these and categorized them based on their purpose (\Cref{fig:clones-interesting-pattern-distribution}). Since some classes served multiple purposes, we assigned multiple categories where appropriate. Visualization was the most common category, appearing in 50\% of clones. This was followed by data processing and statistical analysis.

\subsection{File-Level Clones}
\label{subsec:file-level-clones}

We further investigated \textbf{Notebook Templates} as we saw them often in our qualitative analysis and Jupyter does not offer a structured way handle them. 
Of the \num{1510} Python notebooks and \num{4816} Python scripts, we found 97 classes of file-level clones of which one or more of the files in each class was a notebook. 10 of these classes were identical. We coded the differences in the remaining 87 clones to understand the common patterns of modifications made to the cloned files.

\begin{figure}[b]
    \centering
    \includegraphics[width=\columnwidth]{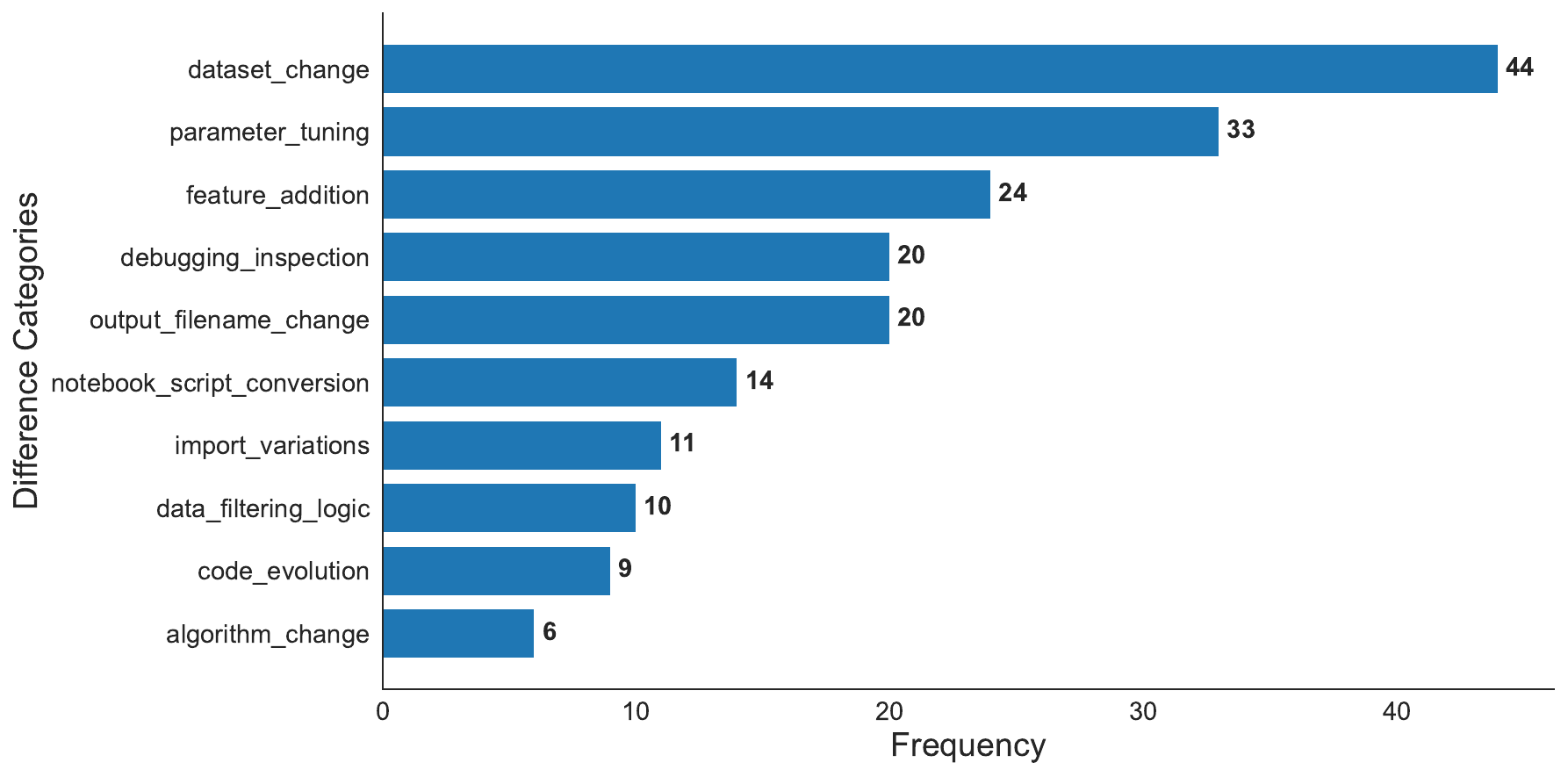}
    \caption{Categories of differences in near-miss file-level clones}
    \label{fig:clones-file-difference-categories}
\end{figure}

\Cref{fig:clones-file-difference-categories} shows the categories of differences we found in the near-miss file-level clones. The most common category was \textbf{dataset change}, where the cloned file was modified to load a different dataset. This was often done by changing the file path or the name of the dataset in the code. The second most common category was \textbf{parameter tuning}, where some parameters in the code were changed to better fit the new dataset or perform a different analysis.

    \section{Variable Lifetime and Mutation}
\label{sec:mutation}

In this section, we analyze two aspects of programs that affect code comprehension: variable lifetimes and mutation. These aspects are indicative of the level of \emph{coupling} in a program, with a high level of coupling reducing reusability of program components. \emph{Variable lifetime} refers to the span over which a variable is used in a program, and \emph{mutation} refers to the modification of data after its initial definition. Longer variable lifetimes separate variable uses from their definitions, potentially making code harder to understand.
Software engineers have argued that frequent use of mutation results in code that is harder to read and maintain~\cite{bloch2008effective}, a claim that has been supported by a controlled study~\cite{10.1109/TSE.2003.1214329}. We analyze both variable lifetimes and mutation and define two metrics, \emph{mutation statement ratio} and \emph{mutation diffusion score} to capture the level of coupling in scientific code.
In understanding coupling in scientific code, we seek opportunities to make such code easier to read and reuse via education, tools, or new languages that encourage locality or purity.

\subsection{Data Sources}

Since both metrics depend on the programming language, we again focused on scripts and Jupyter notebooks written in Python. We used Python scripts and Python-based Jupyter notebooks from our corpus.
To compare the variable lifetime and mutation patterns in scientific code with general Python code, we also collected two additional datasets: one for regular Python scripts and another for non-scientific Jupyter notebooks. The regular Python scripts were collected from DyPyBench~\cite{dypybench}, which contains popular Python libraries that represent typical non-scientific Python development practices. The non-scientific Jupyter notebooks were collected from the dataset by Rule et al.~\cite{rule2018design}, which includes a large number of Jupyter notebooks from public repositories on GitHub. Since dataflow analysis is computationally intensive, we randomly sampled 50 files from each of the four datasets for analysis.

\subsection{Variable Lifetime}

We define \emph{variable lifetime} as the number of statements between a variable's definition and its last use. For example, if a variable is defined at line 10 and last used at line 20, its lifetime is 11 statements. A variable's lifetime is typically smaller than its scope: the last use often comes before the end of the scope. Shorter lifetimes indicate that variables are used in a localized manner, which may promote comprehension because finding a variable's definition and uses becomes easier.

\begin{figure}
    \includegraphics[width=.7\columnwidth]{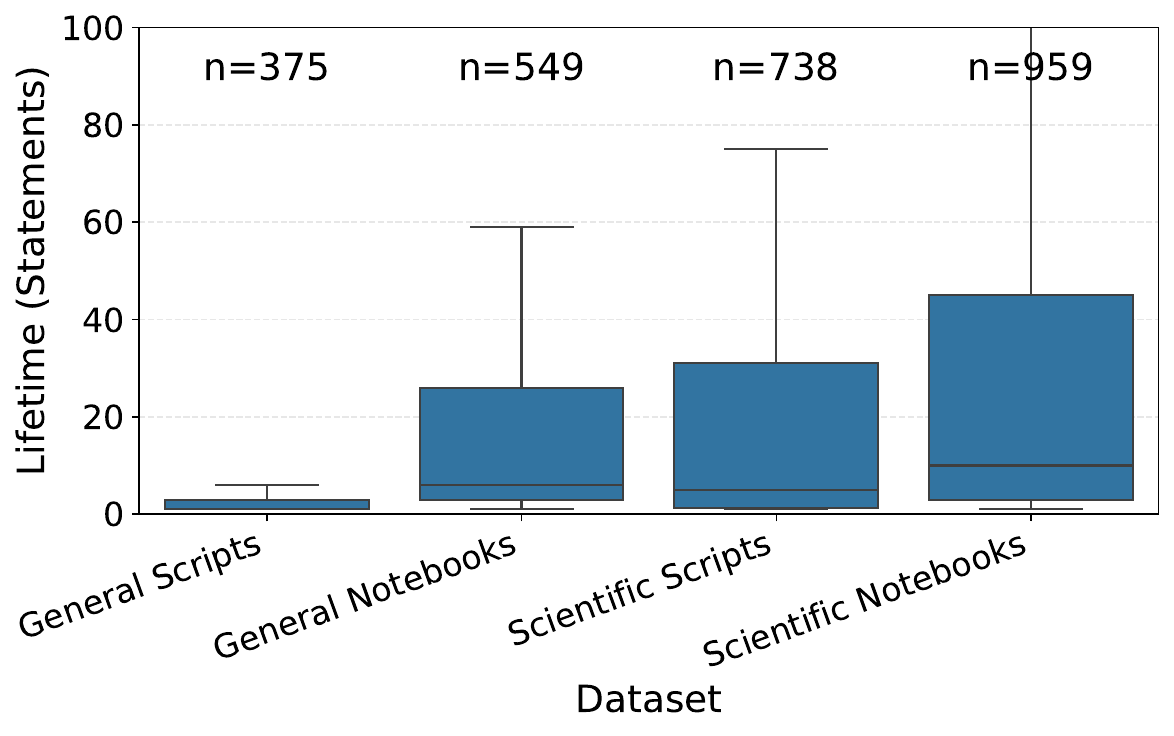}
    \caption{Variable lifetimes in Python scripts and Jupyter notebooks. \texttt{n} indicates the number of variables analyzed in each dataset.}
    \label{fig:lifetime_boxplot}
\end{figure}

\Cref{fig:lifetime_boxplot} shows the computed lifetimes of local variables for each scope (module or function) in each dataset. The average lifetimes were 5 lines in DyPyBench, 20 in Rule et al.'s dataset, 33 for Python scripts in our corpus, and 39 for Jupyter notebooks in our corpus. The extended lifetimes in notebooks (39, compared to 5 in traditional software scripts) suggest that variables persist across larger portions of the code in scientific software than in general-purpose software.

\subsection{Mutation Patterns}

We analyzed mutation patterns in scientific code to understand how variables are defined and subsequently mutated. First, we explain our definition of mutation and how we detect mutations in Python code. Then, we present two metrics to quantify mutations patterns: the \emph{mutating statement ratio} and the \emph{mutation diffusion score}.

\subsubsection{Detecting Mutation}

We define \emph{mutation} as a modification of a variable's value after its declaration. In Python, mutation can happen through augmented assignments (\texttt{x += 1}); subscript assignments (\texttt{lst[i] = value}); attribute assignments (\texttt{obj.attr = value}); and function calls that mutate their arguments (\texttt{lst.append(value)}).
Detecting mutation via assignment can be done syntactically, but determining if a function call mutates its arguments is more complex. We used a dataflow analysis to track variable definitions and mutations. Since Python is dynamically typed and some libraries are not written in Python, not all functions could be analyzed. We relied on two strategies to identify potential mutations: (1) a set of pre-defined rules for common libraries that specify if a function is mutating or not, and (2) a heuristic to classify functions as mutating or non-mutating based on naming conventions.

When we encounter a function call, we first check if the function is defined in the same module. If so, we analyze the function body to determine if it mutates any of its arguments. If the function is imported from another module, we check if it belongs to a set of common libraries for which we have pre-defined specifications (\texttt{json}, \texttt{matlab}, \texttt{numpy}, \texttt{os}, \texttt{pandas}, \texttt{random}, \texttt{shutil}, and \texttt{sklearn}). We chose these libraries because they are widely used in our datasets. If the function is not in our list or the analysis fails to link the function call to its specification, we fall back to name-based heuristics. For example, methods from Python's built-in \texttt{Sequence} type (e.g., \texttt{list}, \texttt{tuple}) with names \texttt{append}, \texttt{extend}, \texttt{insert}, \texttt{remove}, \texttt{pop}, \texttt{clear}, \texttt{sort}, and \texttt{reverse} are classified as mutating, while methods named \texttt{keys}, \texttt{values}, \texttt{items}, \texttt{copy}, and \texttt{get} are non-mutating.
Since our analysis cannot classify all function calls, we report two results for each analysis: an \emph{optimistic} result that assumes unknown function calls do not mutate any of their arguments, and a \emph{conservative} result that assumes unknown function calls mutate all of their arguments.

\subsubsection{Mutating Statement Ratio}
\label{subsubsec:mutating_statement_ratio}

We define the \emph{mutating statement ratio} as the number of mutating statements divided by the total number of statements in a file or notebook. This metric captures how frequently mutations occur in the code---a ratio of 0.2 means that 20\% of all statements modify existing variables rather than creating new ones or performing read-only operations. Higher ratios indicate code with more frequent in-place mutation, which can make data flow harder to understand.

\begin{figure}
    \includegraphics[width=\columnwidth]{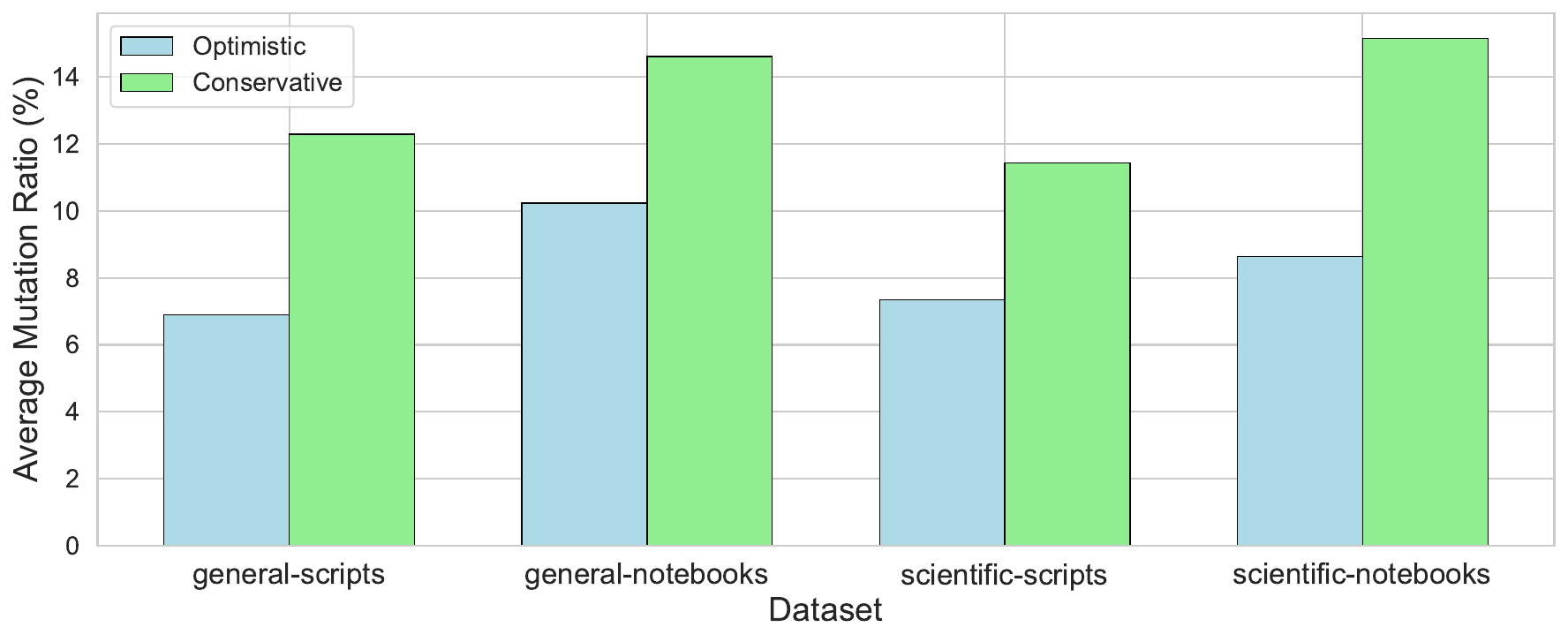}
    \caption{Average mutation ratios in Python scripts and Jupyter notebooks.}
    \label{fig:mutation_ratios}
\end{figure}

\Cref{fig:mutation_ratios} shows the average mutation ratios in functions and modules in our dataset. We observe that the mutation ratios are similar across the datasets. The average optimistic mutation ratios ranged from \num{7}\% to \num{10}\%, while the average conservative mutation ratios ranged from \num{11}\% to \num{15}\%. While the domains and contexts of the datasets differ, they had similar amounts of mutations. However, since variables in scientific code have longer lifetimes, mutations may be spread across many lines of code and make it more challenging to understand and maintain code.

\subsubsection{Mutation Diffusion Score}

The mutating statement ratio does not consider the distance between variable definitions and their subsequent mutations. To address this limitation, we developed the \emph{mutation diffusion score} metric to quantify how distributed mutations are through programs.

Even if two codebases have the same number of mutations and mutation ratio, the distribution of mutations can differ and impact code comprehension. For example, consider the code snippets in \Cref{fig:mutation_patterns}. Both snippets consist of the same seven lines of code, two of which are mutating (the \texttt{append} calls). However, the first snippet groups all mutations together before computing the sum and average, while the second snippet interleaves mutations with other operations. We argue that \Cref{fig:grouped_mutations} is easier to understand because the mutations are localized, while the second snippet requires tracking the state of \texttt{l} across multiple operations.

\begin{figure}[h]
    \centering
    \begin{subfigure}[t]{0.48\columnwidth}
        \begin{minted}[linenos,fontsize=\small]{python}
l = []
l.append(1)  # Mutation 1
l.append(2)  # Mutation 2
s = sum(l)
a = sum(l) / len(l)
m = max(l)
\end{minted}
        \caption{Grouped mutations}
        \label{fig:grouped_mutations}
    \end{subfigure}
    \hfill
    \begin{subfigure}[t]{0.48\columnwidth}
        \begin{minted}[linenos,fontsize=\small]{python}
l = []
s = sum(l)
l.append(1)  # Mutation 1 (+1)
a = sum(l) / len(l)
m = max(l)
l.append(2)  # Mutation 2 (+2)
\end{minted}
        \caption{Scattered mutations}
        \label{fig:scattered_mutations}
    \end{subfigure}
    \caption{Two code snippets with identical mutation count but different mutation patterns. The left snippet groups all mutations together before using the list, while the right snippet interleaves mutations with other operations.}
    \label{fig:mutation_patterns}
\end{figure}

To capture this intuition, we define \emph{mutation diffusion score} using dataflow analysis on a control flow graph. The score quantifies how scattered mutations are by tracking the distance between variable definitions and their subsequent mutations. For each mutation, the score adds the number of lines traversed since the variable was last defined or mutated. For branches, we take the union of all possible paths to account for all potential distances. The mutation diffusion score of a program is the sum of these contributions across all mutations.

In \Cref{fig:grouped_mutations}, the mutation diffusion score is \num{0} because all mutations occur immediately after the variable definition. In contrast, in \Cref{fig:scattered_mutations}, the score is \num{3}. The first mutation \texttt{l.append(1)} on line 3 contributes \num{1} (distance from definition on line 1 through line 2), and the second mutation \texttt{l.append(2)} on line 6 contributes \num{2} (distance from the previous mutation through lines 4 and 5). We provide a formal definition of the metric in Appendix~\ref{app:mutation_diffusion}.

\begin{figure}[b]
    \includegraphics[width=\columnwidth]{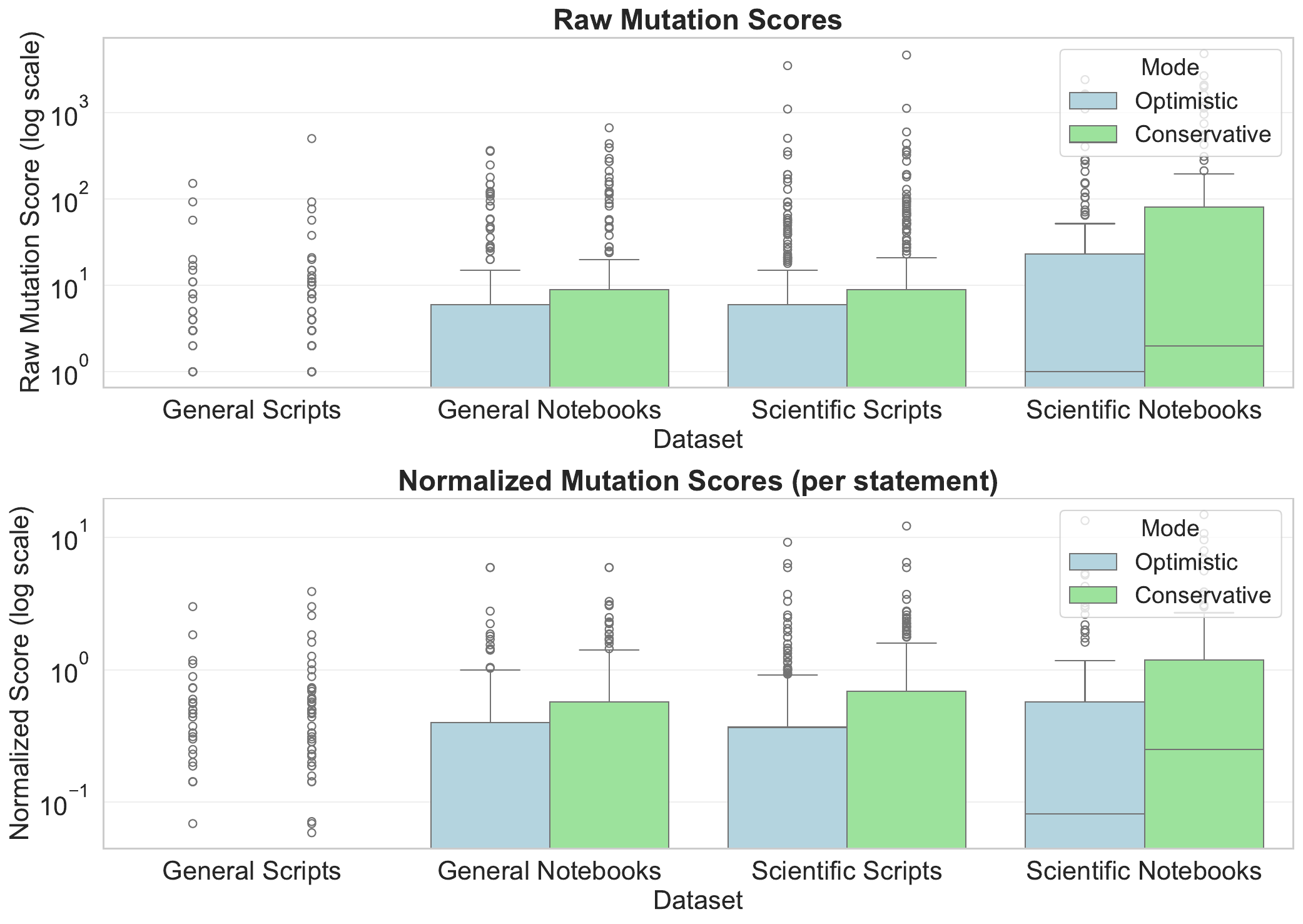}
    \caption{Raw and normalized mutation diffusion scores}
    \label{fig:mutation_score}
\end{figure}

\Cref{fig:mutation_score} shows the distribution of mutation diffusion scores from each dataset, as well as mutation diffusion scores normalized by the number of statements to account for differences in code size, in log scale. On average, notebooks from our corpus had the highest mutation diffusion scores, followed by notebooks from Rule et al.'s dataset, scripts from our corpus, and DyPyBench. This shows that not only do scientific code and notebooks have longer variable lifetimes, but their mutations are more scattered throughout the code. This means that people reading and modifying such code need to understand a large portion of the code to track variable state.

    \section{Discussion}
    \label{sec:discussion}

    Prior work has shown that only 24\% of Jupyter notebooks sampled from Github are fully executable~\cite{pimentel_large-scale_2019}, and our study found a similar story for scientific notebooks. Even with manual troubleshooting, only two of 19 notebooks (11\%) were reproducible. In a sense this is surprising since we would expect that rigorous published work would be associated with better code than randomly selected Github repositories. There are many possible root causes of the low reproducibility we observed. First, scientists lack formal software engineering training so may not be familiar with best practices, especially with regards to managing package and file dependencies. Second, the notebooks in our corpus are larger on average than general notebooks, and this increased complexity may make them more prone to issues. Finally, key aspects of the Jupyter environment, particularly out-of-order cell execution, do not promote reproducible code development. Our subsequent analyses revealed that scientific notebooks are characterized by inconsistent documentation, prevalent code clones, and diffuse mutation, all of which negatively affect their readability and reusability.


    The adoption of guidelines like the FAIR Principles of Scientific Software~\cite{barker_introducing_2022} shows that code quality is important to scientists, but our findings show significant opportunity for improvement. If we want to raise the quality of code published alongside scientific papers, we must design programming systems that promote reproducibility, readability, and reusability, while still supporting the exploratory nature of scientific code development. Jupyter excels as an interactive environment for exploration, but its flat structure, single scope, and out-of-order cell execution make it prone to mess, thus requiring scientists to adopt strategies like rerunning notebooks from the top to avoid unexpected state~\cite{huang_how_2025}. A challenge for future tool designers is to balance flexibility, which promotes exploration, against enforced structure, which promotes code quality. For example, a computational notebook that requires linear cell execution would ensure a program's state is reproducible but would restrict the interactive program flow that users rely on to iterate quickly. Alternatively, reproducible state could be accomplished by disallowing mutation, but this is unsuitable for the imperative programming style that many notebooks use. Identifying nuanced restrictions through careful consideration of tradeoffs is key to future designs.

    \subsection{Documentation}
    Computational notebooks have been considered the laboratory notebooks of scientific computing~\cite{perkel2018jupyter}. Traditional lab notebooks document the scientific process, incorporating formal experimental parameters and the researcher's spontaneous observation into one narrative to record how ideas form and the decision-making process~\cite{higgins2022considerations}. Computational notebooks like Jupyter are meant to provide the same information, allowing the researcher and others to trace the process of discovery by running each step of the computational narrative~\cite{perkel2018jupyter}.

    However, the Jupyter notebooks in our corpus primarily documented low-level code behavior, and not the overarching research process. There is an opportunity for a new computational environment where the finalized code, the decision-making process, and observations co-exist to give a complete record of the scientific process. Almost half of all Markdown cells in scientific notebooks were Headline cells, which are used to structure and divide up a notebook. This suggests that scientists desire more structure in their notebooks, but the flat nature of interspersed Markdown cells limits their ability to incorporate a deeper sense of structure into their notebooks. Providing more avenues for organization in notebooks, such as a designated place for research questions and annotations of results may help scientists more thoroughly document their work. With inspiration from tools such as Variolite and Fork-It~\cite{kery2017variolite, weinman2021forkit}, new interfaces could organize notebooks into a decision-tree of variants. New designs for computational notebooks could mimic and improve upon physical lab notebooks, allowing for clear communication of the scientific process and results.

    \subsection{Clones}
    The prevalence of code cloning in scientific Jupyter notebooks may stem from several factors. In exploratory notebook programming, quick iteration is typically prioritized over code quality~\cite{kery2017variolite}, which may result in more clones since copy-and-pasting code is usually faster (in the short term) than refactoring it. Software engineering concepts like abstraction may be unfamiliar to scientists, as well as relevant technical knowledge, like how to create Python modules. Additionally, code abstracted into a function is not as interactive as code broken across notebook cells~\cite{rawn2025_pagebreaks}, so clones may be the preferable way to reuse code in Jupyter.

    Though there is a maintainability cost to code clones, there is also a cost to abstraction. By the Cognitive Dimensions of Notation~\cite{blackwell_cognitive_dimensions}, abstraction reduces visibility, which is important to notebook programmers~\cite{huang_how_2025}, and requires up-front decision-making that may be premature or impractical for exploratory work. For these reasons, we believe code clones are an inherent, and even advantageous, part of notebook programming. One direction of future work should focus on developing tools that aid clone maintainability, perhaps by tracking clone instances and supporting global edits. Other work may focus on developing notebook-specific forms of abstraction to promote modular design. Our analysis of file-level clones suggests an opportunity for tools that make notebooks reusable. One existing solution is Papermill, a Jupyter extension for parametrizing notebooks~\cite{papermill}. This is a suitable solution for notebooks that are structurally identical, but for notebooks that contain divergence points, a more granular form of abstraction may be appropriate.

    \subsection{Mutation}
    Having many distant mutations in a program requires a user to reason about complex state changes, which is why many software engineering practitioners and researchers recommend limiting state changes as much as possible~\cite{coblenz2016exploring}. The mutation analysis showed that scientific notebooks tend to have many mutations scattered throughout the notebook. When combined with the possibility of out-of-order execution in Jupyter, reasoning about state changes in a notebook becomes even more difficult and inhibits the readability of the code.

    One potential solution is to educate scientists through programs such as Software Carpentry~\cite{softwarecarpentry} to structure their programs in a way that reduces the number of these distant mutations. However, maintaining this discipline can be difficult in the exploratory context many scientists work in. An alternative approach is to create new tools and language designs that enforce specific state management practices. Languages such as Rust require users to distinguish mutable and immutable variables and require the programmer to adhere to its strict ownership and borrowing discipline. While this approach is very successful at preventing a wide range of memory safety bugs, Rust is also considered difficult to learn and use, even by experienced software engineers~\cite{fulton2021benefits}. Thus, it may be necessary to develop new techniques that provide similar correctness guarantees as Rust does while loosening the restrictions Rust puts on its users. One avenue for future work is tailoring a language for the scientific workflow that allows mutation in certain, safe cases automatically and restricts mutation in others. With static and dynamic analyses of Python code in notebooks, we can assess when code is attempting to mutate an aliased reference, and disallow it from running.

    \section{Related Work}
    \label{sec:related-work}

    Here we provide an overview of scientific software engineering studies, as well as relevant work for our four analyses: reproducibility, documentation, code clones, and mutation.

    \subsection{Scientific programming and software engineering}
    Early work identified a growing chasm between science and computing, where software engineering practices were not being developed for or disseminated to domain scientists \cite{kelly_software_2007,faulk_scientific_2009}. More recent research has shown that many software engineering practices are being adopted into scientific contexts, but that challenges still remain for improving productivity, code quality, correctness, and reproducibility \cite{heaton_claims_2015,storer_bridging_2017,arvanitou_software_2021}.
    Others have formalized the characteristics of scientific programming to better distinguish it from software engineering. Segal \cite{segal_problems_2007} classifies scientists as end-user programmers, people for whom software development is means to other professional goals. Kelly \cite{kelly_scientific_2015} argues against this and presents an alternate model of scientific software development that is driven by knowledge acquisition. Pertseva et al. \cite{pertseva_theory_2024} construct a theory of scientific programming efficacy which identifies six contributing factors, including software engineering practices, education, and collaborative atmosphere. Based on software engineering knowledge and the unique requirements of domain science, best practices for scientific programmers have been suggested, with emphases on correctness and reproducibility \cite{dubey_good_2022,lowndes_our_2017,kellogg_role_2019}.

    \subsection{Reproducibility}
    There have been several large-scale studies related to reproducibility of scientific software. Trisovic et al.~\cite{trisovic_large-scale_2022} found that of \num{9078} R files in \num{2109} replication packages only 2\% were executable and 40\% were executable after an automated cleaning phase which involved solving common problems like removing absolute file paths or installing required libraries. Studies of software from specific disciplines have found generally low reproducibility rates, both in executability of code and production of reported results~\cite{tiwari_reproducibility_2021,chang_is_2022,samuel_computational_2024}. Others have assessed the reproducibility of Jupyter notebooks. Pimentel et al.~\cite{pimentel_large-scale_2019} found that only 24\% of Jupyter notebooks were executable and a mere 4\% resulted in the same output. Methods for restoring non-executable notebooks have been proposed, including through static analysis of cell dependencies~\cite{wang_assessing_2020} and LLM-guided fixes~\cite{nguyen_are_2025}, as well as linters that suggest ways to improve reproducibility~\cite{pimentel_understanding_2021}.

    \subsection{Documentation}
    Prior work has analyzed how research code is documented in order to provide recommendations to domain-scientists to help make their code and workflow more understandable. Trisovic et al.~\cite{trisovic_large-scale_2022} found that most authors of research code include documentation alongside their code, including comments or Readme files, but the authors do not compare their findings with non-research code. Domain scientists have also written guidelines for documenting software~\cite{gallagher2024ten,filazzola2022call,osborne2014ten, zielinski2023keep}. Outside of specific recommendations, Hermann and Fehr~\cite{hermann2022documenting} conducted a case study of three large research software projects to understand \emph{why} and \emph{for whom} software is documented. They found that while software is documented, there is a mismatch between the kinds of documentation that helps domain scientists (e.g., why certain decisions were made during the scientific process) and what appears in the code documentation (e.g., explaining implementations of mathematical formulas).
    In our work, we focused on the differences between the scientific notebooks and high-quality data science notebooks, extending an analysis from Wang et al.~\cite{wang2022documentation} to include new categories and analyzing inline comments.

    \subsection{Code clones}
    Code cloning has long been discussed as a bad practice in software engineering~\cite{10.5555/311424} that makes code harder to maintain since changes to clones must be made at each site~\cite{roy_survey_2007,juergens_code_2009,geiger_relation_2006}. However, not all clones are inherently harmful. Kasper et al.~\cite{kapser_cloning_2006} discuss motivations for cloning, including forking, templating, and customization---all patterns we observed in our corpus---and consider the advantages and disadvantages within each context. Other research finds that clones are not as defect-prone as previously thought~\cite{rahman_clones_2012,geiger_relation_2006}. Islam and Zibran~\cite{islam_characteristics_2018}  distinguish between buggy and non-buggy clones and find that buggy clones are significantly larger, more complex, worse documented, and more coupled.

    Two prior studies investigated code cloning in Jupyter. Källén et al.~\cite{kallen_jupyter_2021} analyzed \num{2.7} million Jupyter notebooks from Github and found that over \num{80}\% of snippets are exact or near-miss clones and \num{50}\% of notebooks have no unique code. However, they note that these numbers are likely inflated by tutorials and course materials, where duplication is expected. We found far fewer instances of full notebook clones: only 6.4\% of our files were classified as being file-level clones (97 files), and only \num{0.7}\% were exact duplicates (10 files).

    Koenzen et al.~\cite{koenzen2020_jupyterClones} looked for within-repository clones from a sample of \num{1000} Github repositories and found a mean self-duplication rate of \num{7.6}\%. By manually coding the clones, they found that visualization, machine learning, function definition, data science, and general programming were the top five activities performed by the duplicated code. We performed a similar coding process, and likewise found that visualization was the most common clone activity, followed by data processing, statistical analysis, utility functions, and boilerplate. Since we followed an independent, inductive coding process, we would not expect perfect alignment between the two categorizations. One noteable difference we see between scientific notebooks and general notebooks (where data science is the dominant use) is that machine learning is far more present in general notebooks, while statistical analysis (which Koenzen classifies as mathematics and is the eighth most common activity in their corpus) is more prevalent in science.

    \subsection{Mutation}
    Conventional software engineering wisdom says mutation harms program comprehension~\cite{abelson1985,bloch2008effective}. \citet{10.1109/TSE.2003.1214329} validated this by showing C code containing side-effects was less understandable than its non-mutating counterpart. We used mutation as a measure of notebook coupling. Most coupling metrics rely on the existence of predefined components~\cite{papamichail2019measuring}, which notebooks often lack, or were designed for object-oriented programs~\cite{mehboob_reusability_2021}. Grotov et al.~\cite{grotov_large-scale_2022} defined a coupling metric for notebooks, measuring the average number of shared variables between cells. Our mutation diffusion score measures a more nuanced form of coupling, concerned with hidden dependencies and distance between mutations. We note that this metric needs further study to be validated.

    \section{Conclusion}
    \label{sec:conclusion}


    We set out to understand the extent to which scientific notebooks meet the goals of open science, leveraging a comparative approach relative to general-purpose software to identify opportunities for improvement. Unlike general software, which often uses continuous integration to ensure reproducibility, only two of 19 artifacts reproduced their original results. We found that scientific notebooks are often more sparsely documented, with the authors relying on accompanying papers to provide high-level insight about analyses. Scientific notebooks contain many code clones, suggesting that there may be opportunities for tool designers to help scientists package, find, and reuse relevant code. Finally, scientific notebooks are more \emph{diffuse} than other Python software, with longer variable lifetimes and mutation that is interspersed with non-mutating code.

    On the basis of prior evidence that mutation makes code harder to understand, this suggests that approaches that help scientists organize code could help make scientific code easier to understand. Key opportunities lie in development environments, which could scaffold creation of reusable components; programming languages, which could afford those components and enable scientists to reason more soundly about reuse and composition; and education, which could expose scientists to compositional reasoning principles that have benefited general software engineering practice.

    \section{Data Availability}

    We provide the scripts used to collect and analyze our corpus as supplemental material. The corpus itself is not included, as it contains code that is not licensed for redistribution. However, the repositories are publicly available on Zenodo and GitHub, and we include a program and instructions for obtaining them. We also provide our implementations for computing lifetime and mutation diffusion scores.

    \bibliography{refs,software,sample-base}

    \pagebreak
    \setcounter{page}{1}
    \appendix
    \section{Definition of the Mutation Diffusion Score}
\label{app:mutation_diffusion}

The mutation diffusion score quantifies how scattered mutations are throughout a program by tracking the distance between variable definitions and their subsequent mutations. We define this metric using a dataflow analysis that propagates line number information through the program's control flow graph (CFG). The CFG consists of basic blocks connected by control-flow edges, with each block containing a sequence of statements.

The computation proceeds in two phases:
\begin{enumerate}
    \item \textbf{Dataflow Analysis Phase}: We perform a forward dataflow analysis that tracks, for each variable, the set of line numbers traversed since its last definition or mutation. This analysis iterates until reaching a stable state where the dataflow information no longer changes.
    \item \textbf{Score Calculation Phase}: Using the computed dataflow information, we calculate the mutation diffusion score by summing contributions from each mutation point, where each contribution equals the size of the line number set accumulated for that variable.
\end{enumerate}

\paragraph{Variable Sets}

The analysis distinguishes between two types of variable operations. For each statement $s$, we define two sets:
\begin{itemize}
    \item $\text{DEF}(s)$: variables defined or newly assigned at statement $s$
    \item $\text{UPDATE}(s)$: variables mutated (modified in-place) at statement $s$
\end{itemize}

\paragraph{State Representation} To track mutation patterns, our analysis keeps track of the line numbers where each variable is available in scope but has not been mutated since its last definition or mutation. As a result, our analysis state maps each variable to a set of line numbers:

$$\text{State} = \text{Var} \rightarrow \text{Set}(\mathbb{N})$$

\paragraph{Transfer Functions}

The dataflow analysis propagates line number information through the program. For a statement $s$ at line $\ell$, the transfer function $f_s$ is defined as:
$$f_s(\sigma) = \sigma''$$

With intermediate states:
\begin{align}
    \sigma'  & = \{v \mapsto S \cup \{\ell\} \mid (v, S) \in \sigma\} \quad \text{(add current line to all variables)}                                \\
    \sigma'' & = \sigma'[\text{DEF}(s) \mapsto \emptyset, \text{UPDATE}(s) \mapsto \emptyset] \quad \text{(empty sets for mutated/defined variables)}
\end{align}

Intuitively, the transfer function first adds the current line number $\ell$ to the set of line numbers for all variables in scope. Then, it clears the sets for any variables that are defined or mutated at statement $s$, as these variables have been modified and should not contribute to mutation diffusion.

\paragraph{Phase 1: Computing Dataflow Information}

\subparagraph{Block-Level Analysis}

Within a basic block $B$ containing statements $s_1, s_2, \ldots, s_n$, the analysis iterates through each statement sequentially:
\begin{align}
    \text{IN}[s_1]     & = \text{IN}[B]                                                    \\
    \text{OUT}[s_i]    & = f_{s_i}(\text{IN}[s_i]) \quad \text{for each } i = 1, \ldots, n \\
    \text{IN}[s_{i+1}] & = \text{OUT}[s_i] \quad \text{for each } i = 1, \ldots, n-1
\end{align}

The block-level output is the output of the last statement:
$$\text{OUT}[B] = \text{OUT}[s_n]$$

\subparagraph{Meet Operation}

The analysis must handle control flow joins where multiple execution paths converge. We define the meet operation for combining states as follows:

$$\text{IN}[B] = \bigwedge_{P \in \text{pred}(B)} \text{OUT}[P]$$

Where the meet operation takes the union of sets for each variable:
$$(\sigma_1 \wedge \sigma_2)(v) = \sigma_1(v) \cup \sigma_2(v)$$

\subparagraph{Iterative Computation}

The dataflow analysis begins with an empty initial state at the program's entry point, meaning no variables have accumulated any line numbers yet. It then iteratively applies the transfer functions along all paths in the control flow graph until the dataflow information stabilizes—that is, when further iterations produce no changes to the IN and OUT states for any block in the program.

\paragraph{Phase 2: Computing the Mutation Diffusion Score}

After the dataflow information has stabilized, we calculate the mutation diffusion score by examining each mutation point in the program. For each mutation, we determine its contribution based on the dataflow state at that point. Formally, for a statement $s_i$ in block $B$ and each variable $v \in \text{UPDATE}(s_i)$, the contribution is:
$$\text{Contrib}(s_i, v) = \begin{cases}
        |\sigma_{\text{IN}[s_i]}(v)| & \text{if } v \in \text{UPDATE}(s_i) \\
        0                            & \text{otherwise}
    \end{cases}$$

This contribution represents the "distance" the variable has traveled (in terms of lines executed) since it was last defined or mutated. The total mutation diffusion score is the sum of all such contributions across the entire program:
$$\text{Score} = \sum_{B \in \text{CFG}} \sum_{i: s_i \in B} \sum_{v \in \text{UPDATE}(s_i)} \text{Contrib}(s_i, v)$$

\end{sloppypar}

\end{document}